\documentstyle[aps,preprint,epsf,aps]{revtex}
\begin{document}
\draft
\title{Induced Pseudoscalar Coupling Constant}
\author{ Il-Tong Cheon 
\footnote{e.mail : itcheon@phya.yonsei.ac.kr} and Myung Ki Cheoun 
\footnote{e.mail : cheoun@phya.yonsei.ac.kr} }
\address{
 Department of Physics, Yonsei University,
Seoul, 120-749, Korea \\ (9 November, 1998)}
\maketitle
\begin{abstract}
The recent TRIUMF experiment for $\mu^- p \rightarrow n \nu_{\mu} 
\gamma$ gave
a surprising result that the induced pseudoscalar coupling constant
$g_P$ was larger than the value obtained from
$\mu^- p \rightarrow n \nu_{\mu}$ experiment as much as 44 \%.
Reexamining contribution of the axial vector current,
we found an additional term to
the matrix element of Beder and Fearing which was used
to extract the $g_P$ value from the measured photon
energy spectrum. This additional term plays a key role to restore the
reliability of $g_P ( - 0.88 m_{\mu}^2 ) = 6.77 g_A (0)$.
\end{abstract}
\vspace{1cm}

In semileptonic weak interaction, the strong force can generally
induce four couplings additional to
the usual vector and axial vector couplings, i.e. weak magnetic 
$G_M$, pseudoscalar $G_P$, scalar $G_S$ and tensor $G_T$.

The matrix element of vector and axial vector currents
are given as
\begin{eqnarray}
\langle N (p^{'}) \vert V_a^{\mu} (0) \vert N (p) \rangle &=
{\bar u} ( p^{'}) [ G_V ( q^2) \gamma^{\mu} + {{G_S ( q^2 )} \over { 2 m}}
q^{\mu} + G_M ( q^2) \sigma^{\mu \nu} q_{\nu} ] {\tau_a \over 2} u(p)
  \\ \nonumber
\langle N (p^{'}) \vert A_a^{\mu} (0) \vert N (p) \rangle &=
{\bar u} ( p^{'}) [ G_A ( q^2) \gamma^{\mu} + {{G_P ( q^2 )} \over { 2 m}}
q^{\mu} + G_T ( q^2) \sigma^{\mu \nu} q_{\nu} ] \gamma_5 
{\tau_a \over 2} u(p)~,
\end{eqnarray}
where $G_A (0) = g_A (0),~ G_M(0) = g_M(0),~ G_V(0) = g_V(0)$ and 
$~G_P ( q^2) = ( {{ 2 m } \over { m_{\mu}}} )
g_P ( q^2) $ with the nucleon and muon masses, $m $ and $ m_{\mu}$. 
$\tau_a$ is the isospin operator. $G_S$ and $G_T$ belong to the
second class current which has a different G-parity from the first
class current, and they are assumed to be absent from the
muon capture to be discussed in this paper. 
On the basis of the PCAC (Partially
Conserved Axial Current), the induced pseudoscalar coupling constant is
calculated as
\begin{equation}
g_P ( -0.88 m_{\mu}^2 ) = { { 2 m ~  m_{\mu} } \over 
{ m_{\pi}^2 + 0.88 m_{\mu}^2}} g_A (0) = 6.77 g_A(0 ).
\end{equation}
This value is confirmed by an experiment of the 
ordinary muon capture (OMC) on a proton, ${\mu}^- p \rightarrow
n \nu_{\mu}$ \cite{Ba81}.

However, such kind of determination of $g_P$ value induces 25 \%
uncertainty at least, because the momentum transfer is far from 
the pion pole. It is extremely important to obtain a precise value
for $g_P$ of the weak hadronic current, because
it plays a key role in the fundamental weak interaction
processes. The only way to approach to the pion pole is
the radiative muon capture (RMC) on a proton, $\mu^- p \rightarrow
n \nu_{\mu} \gamma$.

Recently, the TRIUMF group measured the RMC photon energy
spectrum and extracted a surprising result \cite{Jo96}
\begin{equation}
{\hat g_P} \equiv g_P ( - 0.88 m_{\mu}^2 ) / g_A (0) = 9.8 \pm 0.7 \pm 0.3~.
\end{equation}
It exceeds the value obtained from the OMC as much as 44\%. 
This discrepancy is serious because the theoretical value of
$g_P$ is predicted in a fundamental manner
based on the PCAC imposed on the axial vector current and
agrees with the OMC value.
As long as the PCAC is assumed to be creditable, a doubt may be
cast on the result of TRIUMF experiment. However, the measured
photon energy spectrum seems to be reliable in view point of their
enough experimental experiences in TRIUMF. In order to solve this puzzle,
one has to reexamine carefully the Beder-Fearing formula \cite{Fe80,Be87}
used to extract the $g_P$ value from the measured spectrum.

The chiral perturbation calculation (ChPT) was recently
carried out for OMC \cite{Fe97} and well reproduced the
PCAC prediction, i.e. ${\hat g}_P$ = 6.77. It is also
consistent with the result of the heavy baryon chiral
perturbation theory \cite{Be94}. Therefore, we
believe that the calculation based on PCAC is reliable. Since
the RMC amplitude is generated merely
by the minimal coupling procedures from the OMC amplitude \cite{Fe80},
the calculation method based on PCAC might preserve the confidence 
even for the RMC.

The axial current coupled to the electromagnetic field can be 
derived in an elegant manner, 
i.e. a standard gauge transformation. We start from the ordinary linear 
$\sigma$-model given by the Lagrangian
\begin{equation}
{\cal L}_0 = {\bar \Psi} [ i \gamma^{\mu} \partial_{\mu} +
g ( \sigma + i {\vec \tau} \cdot {\vec \pi} \gamma_5 )] \Psi
+ { 1 \over 2} [ {( \partial_{\mu} {\vec \pi} )}^2 +
{( \partial_{\mu}{\sigma} )}^2 ]
 - { 1 \over 2} {\mu}^2 ( {\vec \pi}^2 + {\sigma}^2 )
- { {\lambda} \over 4} {( {\vec \pi}^2 + {\sigma}^2 )}^2~.
\end{equation}
Although this Lagrangian is invariant under the SU(2) isovector
infinitesimal chiral and gauge transformations, 
$ \Psi \rightarrow {\Psi}^{'} =  ( 1 + i {\gamma}_5 {\vec \eta} \cdot
{{\vec \tau} \over 2} ) \Psi , 
{\bar \Psi} \rightarrow {{\bar \Psi}}^{'} 
= {\bar \Psi} ( 1 + i {\gamma}_5 {\vec \eta} \cdot
{{\vec \tau} \over 2} ) , $ $\sigma \rightarrow {\sigma}^{'} = 
\sigma + {\vec \eta} \cdot {\vec \pi}$ and ${\vec \pi} \rightarrow
{\vec \pi}^{'} = {\vec \pi} - {\vec \eta} \sigma$, it describes only a
massless fermion. In order to create the pion mass, 
the chiral symmetry breaking term $\zeta\sigma $ 
should be included into ${\cal L}_0$. 
Since the
vacuum expectation value of the $\sigma$ field does not vanish,
i.e. $\langle 0 \vert \sigma \vert 0 \rangle =  f_{\pi}$, 
the $\sigma$ field is shifted as $\sigma \rightarrow {\tilde \sigma} =
\sigma - f_{\pi}$ and, then, $ \langle 0 \vert {\tilde \sigma} \vert 0 \rangle$
 = 0 is fulfilled. Then, the pion and sigma meson's masses can be given as
$m_{\pi}^2 = {\mu}^2 + \lambda f_{\pi}^2 $, 
$m_{\sigma}^2 = {\mu}^2 + 3 \lambda f_{\pi}^2$ and 
$\zeta = f_{\pi} m_{\pi}^2$, and the nucleon mass is $m = - g f_{\pi}$.

The resulting Lagrangian without the symmetry breaking term 
$\zeta\sigma $ becomes
\begin{eqnarray}
{\cal L}_0 &=& {\bar \Psi} [ i \gamma^{\mu} \partial_{\mu} - m +
g ( {\tilde \sigma} + i {\vec \tau} \cdot {\vec \pi} \gamma_5 )] \Psi
 + { 1 \over 2} [ {( \partial_{\mu} {\vec \pi} )}^2 +
{( \partial_{\mu}{{{\tilde \sigma}}} )}^2 ] \nonumber \\
& &  - { {m_{\pi}^2} \over {2}} {\vec \pi}^2
 - { {m_{\sigma}^2} \over {2}} {\tilde \sigma}^2
 - f_{\pi} [ \lambda ( {\vec \pi}^2 + {\tilde \sigma}^2 ) + m_{\pi}^2]
{\tilde \sigma} 
- { {\lambda} \over 4} {( {\vec \pi}^2 + {{\tilde \sigma}}^2 )}^2 + const~.
\end{eqnarray}
By the relation
\begin{equation}
exp ( { i \over {f_{\pi}}} {\gamma}_5 {\vec \tau} \cdot {\vec \phi})
 = cos ( { \phi \over {f_{\pi}}} ) + i {\gamma}_5 {\vec \tau } \cdot
{\hat \phi} sin ( { \phi \over { f_{\pi}}} ) \equiv { 1 \over { f_{\pi}}}
[ {\tilde \sigma} + i {\gamma}_5 {\vec \tau} \cdot {\vec \pi} ] ~,
\end{equation}
where ${\hat \phi} = {\vec \phi} / {\phi}$, and replacement of
${\tilde \sigma} \rightarrow {\tilde \sigma}^{'} = { m \over g} +
f_{\pi} cos ( { \phi \over f_{\pi}})$, the Lagrangian can be rewritten
for $g = - m / f_{\pi}$ as
\begin{equation}
{\cal L}_0^{'}  =  {\bar \Psi} [ i \gamma^{\mu} \partial_{\mu} {} +
g f_{\pi} exp (  { i \over {f_{\pi}}} 
{\vec \tau} \cdot {\vec \phi} \gamma_5 )] \Psi
+ { 1 \over 2} {( {\partial}_{\mu} {\vec \phi} )}^2 ~~~,
\end{equation}
where ${\cal L}_0^{'} = {\cal L}_0 - const $.

It is easy to see that the Lagrangian ${\cal L}_0^{'}$ holds a global
chiral symmetry under the infinitesimal phase transformations
${\Psi} \rightarrow {\Psi}^{'} = ( 1 + i \gamma_5 {{\vec \tau} \over 2}
\cdot {\vec \eta} ) \Psi$, ${\vec \phi} \rightarrow {\vec \phi}^{'}
= {\vec \phi} - f_{\pi} {\vec \eta}$. Even when the derivative is
replaced by a covariant derivative,
i.e. $\partial_{\mu} \rightarrow D_{\mu} = \partial_{\mu}
- i e \epsilon_{\mu}$ where $\epsilon_{\mu}$ is the photon
polarization vector, ${\cal L}_0^{'}$ remains in the global chiral symmetry,
since higher order terms than $e^2$ and ${\vec \eta}^2$ are ignored 
and thus, $e {\vec \eta} $ yields negligible
small quantities.

For deriving the axial current from the Lagrangian, let us first
obtain the extended Euler equation by defining
the action,
\begin{equation}
{\cal S} = \int d^4 x {\cal L} ( \phi_i (x) , D_{\mu}^{\pm} \phi_i (x) ) ~,
\end{equation}
where $D_{\mu}^{\pm} = \partial_{\mu} \pm \kappa_{\mu}$.
Under an infinitesimal variation $\delta \phi_i (x)$, the stationary
condition yields
\begin{equation}
0 = \delta {\cal S} = \int d^4 x [ { {\partial {\cal L}  } \over { \partial
\phi_i }} \delta \phi_i + { {\partial {\cal L}  } \over { \partial
(D_{\mu}^{\pm} \phi_i ) }} \delta ( D_{\mu}^{\pm} \phi_i ) ]~.
\end{equation}
Since $\delta ( D_{\mu}^{\pm} \phi_i ) = \delta ( \partial_{\mu} \phi_i
\pm \kappa_{\mu} \phi_i ) = \partial_{\mu} ( \delta \phi_i ) \pm \kappa_{\mu}
\delta \phi_i$, the integrand becomes
\begin{eqnarray}
& & { {\partial {\cal L}  } \over { \partial
\phi_i }} \delta \phi_i + { {\partial {\cal L}  } \over { \partial
(D_{\mu}^{\pm} \phi_i ) }}  ( \partial_{\mu} \delta  \phi_i )
\pm \kappa_{\mu} { {\partial {\cal L}  } \over { \partial
(D_{\mu}^{\pm} \phi_i ) }}  \delta \phi_i  \nonumber \\
& &  =
[ { {\partial {\cal L}  } \over { \partial
\phi_i }} - (\partial_{\mu} \mp \kappa_{\mu} )
{ {\partial {\cal L}  } \over { \partial
(D_{\mu}^{\pm} \phi_i ) }} ] \delta \phi_i
 +    \partial_{\mu} [ { {\partial {\cal L}  } \over { \partial
(D_{\mu}^{\pm} \phi_i ) }} \delta \phi_i ] ~.
\end{eqnarray}
Thus, the extended Euler equation is found as 
\begin{equation}
{ {\partial {\cal L}} \over { \partial \phi_i}}
- D_{\mu}^{\mp} { {\partial {\cal L}} \over { \partial ( D_{\mu}^{\pm} 
\phi_i})} = 0 ~.
\end{equation}
Next, we derive the extended Gell-Mann-Levy equations. The
variation $\delta \phi_i = - \eta_{a} (x) F_i^{a} $ changes 
the Lagrangian ${\cal L}$ as $ {\cal L } + {\delta} {\cal L}$, where
$\delta {\cal L} ( \phi_i , D_{\mu}^{\pm} \phi_i )$. Then,
we have
\begin{eqnarray}
\delta {\cal L} & = &
{{\partial {\cal L}} \over {\partial \phi_i   }} \delta \phi_i +
{{\partial {\cal L}} \over {\partial (D_{\mu}^{(\pm)} \phi_i )  }} 
\delta ( D_{\mu}^{\pm} \phi_i ) \nonumber \\
& = &
- {{\partial {\cal L}} \over {\partial \phi_i   }} \eta_{a} (x) F_i^{a} 
-
{{\partial {\cal L}} \over {\partial (D_{\mu}^{(\pm)} \phi_i )  }} 
( D_{\mu}^{\pm} \eta_{a} (x) ) F_i^{a} \nonumber \\
& = &
- [ D_{\mu}^{\mp} ( {{\partial 
{\cal L}} \over {\partial (D_{\mu}^{(\pm)} \phi_i )  }} 
 F_i^{a} ) ] \eta_{a} (x) -  
[ {{\partial {\cal L}} \over {\partial (D_{\mu}^{(\pm)} \phi_i )  }} 
F_i^{a}] ( D_{\mu}^{\pm} \eta_{a} (x) ) ~, 
\end{eqnarray}
where we used the extended Euler equation in (11). Defining
$(  {   {\partial {\cal L}   } \over {
\partial ( D_{\mu}^{\pm} \phi_i  )  }} F_i^{a}) =  A_{a}^{\mu}$
, we obtain the extended Gell-Mann-Levy equations,
\begin{equation}
A_{a}^{\mu} =  - {   {\partial \delta {\cal L}   } \over {
\partial ( D_{\mu}^{\pm} \eta_{a}  )  }} ~~~,~~~  
D_{\mu}^{\mp} A_{a}^{\mu} = - { { \partial \delta {\cal L}} \over 
{ \partial \eta_{a}}} ~.
\end{equation}

If $\partial_{\mu} \eta_a $ = 0, i.e. $\eta_a$ is independent of $x$,
eq.(12) reduces to $\delta {\cal L} = - ( \partial_{\mu} A_a^{\mu} ) \eta_a$.
Since $\delta {\cal L}$ = 0 if the
lagrangian holds a global symmetry under
the infinitesimal variation, we have $\partial_{\mu} A_a^{\mu} $ = 0.
This statement is synonymous with the Noether's theorem.

Under the local transformations,
${\bar \Psi} \rightarrow {\bar \Psi}^{'} = {\bar \Psi} ( 1 + i \gamma_5
{{\vec \tau } \over 2} \cdot {\vec \eta} (x)  ),
{ \Psi} \rightarrow { \Psi}^{'} = ( 1 + i \gamma_5 {{\vec \tau} \over 2}
\cdot {\vec \eta} (x) ) \Psi,  
{\vec \phi} \rightarrow {\vec \phi}^{'} = {\vec \phi} - f_{\pi}
{\vec \eta} (x) ,$ and  
$\partial_{\mu} \rightarrow D_{\mu} = \partial_{\mu}
- i e \epsilon_{\mu}$, the Lagrangian, ${\cal L}_0^{'}$ reads as
\begin{equation}
{\tilde {\cal L}}_0 =
{\bar \Psi} [  i \gamma^{\mu} D_{\mu} + g f_{\pi}
 exp ( { i \over { f_{\pi}} } \gamma_5 {\vec \tau } \cdot {\vec \phi} ) ]
\Psi
- {\bar \Psi} \gamma^{\mu} \gamma_5 { {\vec \tau} \over 2} \Psi \cdot
(  D_{\mu} {\vec \eta} ) 
+  { 1 \over 2} {(D_{\mu}  {\vec \phi}  )}^2
- (D^{ \mu}  {\vec \phi}  ) \cdot f_{\pi} 
(D_{\mu}  {\vec \eta}  ) ~.
\end{equation}
For this case, eq.(12) yields
\begin{equation}
\delta {\tilde {\cal L}}_0 =
- {\bar \Psi} \gamma^{\mu} \gamma_5 { {\vec \tau} \over 2} \Psi \cdot
(  D_{\mu} {\vec \eta} ) 
- (D^{ \mu}  {\vec \phi}  ) \cdot f_{\pi} 
(D_{\mu}  {\vec \eta}  ) ~.
\end{equation}
Thus, by eq.(13), the axial current and its divergence are given
as
\begin{equation}
A_{a}^{\mu} = {\bar \Psi} \gamma^{\mu} \gamma_5 { \tau_{a} \over 2}
\Psi + f_{\pi} D^{ \mu} \phi_{a}~~,~~ D_{\mu}^{ (+)} 
A_{a}^{\mu} = 0  ~.
\end{equation}
Here, notice that ${D_{\mu}}^{(+)} = \partial_{\mu} + i e \epsilon_{\mu}$
because ${D_{\mu}}^{(-)} = D_{\mu} = 
\partial_{\mu} - i e \epsilon_{\mu}$.
If we add the chiral symmetry breaking
term $ - {   {m_{\pi}^2 } \over {2  }} {\vec \phi}^2 $ to ${\cal L}_0^{'}$,
the second equation in (16) yields
\begin{equation}
D_{\mu}^{ (+)} A_a^{\mu} = - m_{\pi}^2 f_{\pi} \phi_a ~,
\end{equation}
which is the extended form of the PCAC. The
same equation was also given by Adler \cite{Ad65}.
As is seen in eq.(16), the axial coupling constant appears
to be unity. However, it is well known as $g_A = 1.25$. To cure
this defect, the following chiral invariant Lagrangian
is added to ${\cal L}_0^{'}$ \cite{Akh89},
\begin{equation}
{\cal L}_1 = C_1 {\bar \Psi} \gamma^{\mu} \gamma_5 { {\vec \tau} \over 2}
\Psi \cdot f_{\pi} \partial_{\mu} {\vec \phi} ~,
\end{equation}
where $C_1$ is determined so as to give $g_A = 1.25$. Then, the axial current
becomes
\begin{equation}
A_a^{\mu} = g_A {\bar \Psi} \gamma^{\mu} \gamma_5 { {\tau_a} \over 2}
\Psi +  f_{\pi} D^{\mu} {\phi_a} ~,
\end{equation}
where $g_A = 1 + C_1 f_{\pi}^2$.

Operating a covariant derivative ${D_{\mu}^{ (+) }}$ on eq.(19) 
and equating it to eq.(17), we find
\begin{equation}
- f_{\pi} ( D_{\mu}^{ (+)} D^{ \mu} + m_{\pi}^2 ) \phi_a
= g_A D_{\mu}^{ (+) } 
{\bar \Psi} \gamma^{\mu} \gamma_5 { {\tau_a} \over 2}
\Psi ~,
\end{equation}
i.e., in ignoring $e^2$-order term,
\begin{equation}
f_{\pi} ( q_{\mu}^2 - m_{\pi}^2 ) \phi_a
= g_A [ 2 m i {\bar \Psi}  \gamma_5 { { \tau}_a \over 2}
\Psi + i e  
{\bar \Psi} \epsilon_{\mu} \gamma^{\mu} \gamma_5 { {\tau_a} \over 2}
\Psi ] ~.
\end{equation}
When the solution of this equation $\phi_a$ with a definition, $ g_A /
( q^2 - m_{\pi}^2 ) = - g_P / 2 m m_{\mu} $, is substituted into
eq.(19), we obtain
\begin{eqnarray}
 A_a^{\mu} (x)
& = & {\bar \Psi} (x) [ g_A \gamma^{\mu} \gamma_5 +
{  {g_P (q^2)  } \over { m_{\mu}}} q^{\mu} \gamma_5
- {  {e g_P ( q^2)  } \over { m_{\mu}}} \epsilon^{\mu} \gamma_5 ]
{\tau_a \over 2} \Psi (x)  \nonumber \\
& & + {  {e g_P (q^2)  } \over { 2 m m_{\mu}}} q^{\mu}
[ {\bar \Psi} (x) \epsilon_{a} \gamma^{\alpha} \gamma_5
{\tau_a \over 2} \Psi (x) ]~,
\end{eqnarray}
where $e^2$ term is ignored.
The fourth term is
missing \cite{Ch97} in the previous calculations \cite{Fe80,Be87}. 
As will be shown below, 
it is actually a very important term corresponding to the seagull term 
in the pion photoproduction \cite{Be67}.

It may be possible to introduce another chiral invariant
Lagrangian of the form \cite{Akh89}
\begin{equation}
{\cal L}_2 = C_2 {\bar \Psi} \gamma^{\mu} { {\vec \tau} \over 2} \Psi
\cdot ( {\vec \phi} \times \partial_{\mu} {\vec \phi} ) ~,
\end{equation}
but this one does not contribute to the axial current. This fact can be easily 
seen as shown below. When eq.(23) is added to the Lagrangian, the meson
field is obtained by means used above in the
following form instead of eq.(20),
\begin{equation}
{\vec \phi} = {[ 1 + f_{\pi}^2 C_2^2 B^2 ]}^{-1}
[ g_A {\vec B} + f_{\pi} C_2 g_A  ( {\vec B} \times {\vec B} )
 + f_{\pi}^2 C_2^2 g_A {\vec B} ( {\vec B} \cdot {\vec B} ) ] ~,
\end{equation}
where ${\vec B} =  - {  {g_P  }  \over 
{ 2 m m_{\mu} g_A f_{\pi} }} D_{\mu}^{\pi (+) } (  {\bar \Psi}
\gamma^{\mu} \gamma_5 { {\vec \tau} \over 2} \Psi ) $. Since
the second term in eq.(24)
vanishes, it reduces to ${\vec \phi} = g_A {\vec B}$. This is exactly the
same as that given in eq.(21).

Following the Fearing's formulation and notation \cite{Fe80} for diagrams 
given in Fig.1, one can evaluate the relativistic amplitude of RMC
on a proton as
\begin{equation}
M_{ f i} =
M_a + M_b + M_c + M_d + M_e + \Delta M_e
\end{equation}
with
\begin{eqnarray}
M_a &=& - \epsilon_{\alpha} {\bar u}_n \Gamma^{\delta} ( Q) u_p
             \cdot {\bar u}_{\nu} \gamma_{\delta} ( 1 - \gamma_5)
  { { \mu\!\!\!/ - {k \!\!\!/} + m_{\mu}  } \over { - 2 k \cdot \mu}} 
\gamma^{\alpha} u_{\mu} ~,\\ \nonumber
M_b &=&  \epsilon_{\alpha} L_{\delta} {\bar u}_n \Gamma^{\delta} ( K)
{  {{p \!\!\!/}-{ k \!\!\!/}+ m_p} \over {- 2 k \cdot p}} 
(\gamma^{\alpha} - i \kappa_p {  \sigma^{\alpha \beta} \over { 2 m_p}} 
k_{\beta} ) u_p ~, \\ \nonumber
M_c &=&  \epsilon_{\alpha} L_{\delta} {\bar u}_n 
( - i \kappa_n { \sigma^{\alpha \beta} \over {2 m_n  }} k_{\beta} )
{  {{n \!\!\!/}+{ k \!\!\!/}+ m_n} \over { 2 k \cdot n}} 
 \Gamma^{\delta} ( K)  u_p ~,\\ \nonumber
M_d &=&  - \epsilon_{\alpha} L_{\delta} {\bar u}_n 
( { {2 Q^{\alpha}  + k^{\alpha}} \over { Q^2 - m_{\pi}^2 }} 
{ g_P ( K^2)  \over m_{\mu} }
  K^{\delta} \gamma_5 ) u_p ~,\\ \nonumber
M_e &=&   \epsilon_{\alpha} L_{\delta} {\bar u}_n 
( {   {i g_M } \over {2 m}  } \sigma^{\delta \alpha}
 + {{g_P ( Q^2)}  \over{m_{\mu}} } 
\gamma_5 g^{\delta \alpha} )  
 u_p ~,\\ \nonumber
\Delta M_e & =&  - \epsilon_{\alpha} L_{\delta} {\bar u}_n 
( {   { g_P (K^2)} \over {2 m  m_{\mu}}  }  Q^{\delta} \gamma_5 
\gamma^{\alpha} ) u_p ~,
\end{eqnarray}
where 
\begin{equation}
\Gamma^{\delta} ( q ) = g_V \gamma^{\delta} + { { i g_M } \over { 2 m}}
\sigma^{\delta \beta } q_{\beta} + g_A \gamma^{\delta} \gamma_5 +
{  {g_P (q^2)} \over m_{\mu}} q^{\delta} \gamma_5 ~,
\end{equation}
$L_{\delta} = {\bar u}_{\nu} \gamma_{\delta} ( 1 - \gamma_5) u_{\mu},
K = n - p + k $ and $ Q = n - p$ with momenta of neutron,
proton and photon, $ n, p $ and $ k$, respectively. And $m \sim m_p 
\sim m_n$. Other constants are taken as $g_V = 1.0, g_A = - 1.25, g_M = 3.71,
\kappa_p = 1.79$ and $\kappa_n = - 1.91$ \cite{Fe80}. $M_e$ term is
originated from the third term in eq.(22) and 
$\Delta M_e$ term comes from the fourth term. 
But the latter, $\Delta M_e$, is missing in the paper by Fearing
\cite{Fe80,Be87}. Accordingly, this term was not included in
the previous procedure of extracting $g_P$ value from the experimental photon
energy spectrum \cite{Jo96}. The transition rate is given by
\begin{equation}
{ {d \Gamma_{RMC}} \over {d k}} =
{  {\alpha G^2 \vert \phi_{\mu} \vert^2 m_N  } \over {  {( 2 \pi )}^2 }}
 \int_{-1}^{1}  dy 
{  { k E_{\nu}^2  } \over { W_0 - k( 1 - y) }} { 1 \over 4} \sum_{spins}
 \vert M_{f i } {\vert}^{2} ~,
\end{equation}
where $\alpha$ is the fine structure constant, $G$ 
is the standard weak coupling constant, $ y = {\hat k } \cdot 
{\hat \nu},~ k_{max} = ( W_0^2 - m_n^2 ) / 2 W_0,~ E_{\nu} = W_0
( k_{max} - k ) / [ W_0 - k( 1 - \nu)],~ W_0 = m_p + m_n - $
(muon binding energy) and $\vert \phi_{\mu} \vert^2$ is the absolute square
of muon wave function averaged over the proton which
is taken as a point Coulomb. 

In order to compare to the experimental results, we take the following 
steps. For liquid hydrogen target, muon capture is
dominated through the ortho and para $p \mu  p$ molecular states
\cite{Jo96,Ba82}. Since these molecular states can be attributed to the
combinations of hyperfine states of $\mu  p$ atomic states \cite{Ba82}
i.e. single and triplet states, we decompose the statistical spin
mixture ${1 \over 4} \sum_{spins} \vert M_{ f i } {\vert}^2$ into
such hyperfine states by reducing $4 \times 4 $ matrix elements to
$2 \times 2$ spin matrix elements. At this step, we
confirmed that when the $\Delta M_e$ term was not
included, eq.(28) reproduced the curves given in ref. \cite{Be87}. 
Finally, by 
exploiting the mixture of muonic states
relevant in experiments \cite{Jo96}, we calculate the photon energy spectrum. 
The count number of the photons is now expressed as
\begin{equation}
N = Z {  { d \Gamma_{RMC}  } \over {d k  }} ~.
\end{equation}
Here $Z$ is determined by adjusting the value of
$d \Gamma_{RMC} / d k $ without $\Delta M_e$ term for
${\hat g}_P \equiv g_P ( - 0.88 m_{\mu}^2 ) / g_A (0) = 9.8$ so
as to agree with the best fit curve in ref. \cite{Jo96}.
 With this value of $Z$, we have to examine the case of $\Delta M_e$ included.

Our results are shown in Fig.2. The solid curve is for the spectrum obtained
in ref. \cite{Jo96}, i.e. the result without $\Delta M_e$ term for 
${\hat g}_P = 9.8$. On the other hand, the dotted curve is
calculated without $\Delta M_e$ term for ${\hat g}_P$ = 6.77. This curve 
is obviously much lower than the measured spectrum. When 
$\Delta M_e$ term is taken into account for ${\hat g}_P$ = 6.77, we obtain
the dashed curve which is very close to the solid curve. The minor discrepancy
may be due to the neglect of higher order contribution and
other degree of freedom such as $\Delta$. Our result shows that $\Delta M_e$
term restores the credit of ${\hat g}_P = 6.77$.

The number of RMC photons observed for $k \ge 60 MeV $ is 279 $\pm$ 26 and
the number of those from the solid curve is 299, while 
our result obtained by integrating the dotted curve
spectrum is 273. Since the contribution of $\Delta$ degree of freedom
is known to be a few percent \cite{Be87}, it is not included in the present
calculation. Vector mesons such as $\rho$ and $\omega$ make also very
small contributions. Higher order terms are
pointed out to be insignificant \cite{Fe80}.

The pion field actually interacts in virtual state with the nucleon
and therefore the $\pi N N$ form factor may be taken into account
as an off-shell effect. However, the standard $\pi N N$ form factor
$f_{\pi N N} ( q^2) = ( \Lambda^2 - m_{\pi}^2 ) / ( \Lambda^2 - q^2)$
with $\Lambda^2 = m_{\rho}^2 + m_{\pi}^2 $ participates through 
an effective $g_P ( q^2 )$, i.e. 
${\tilde g}_P ( q^2 ) = g_P (q^2) f_{\pi N N} ( q^2 )$ but
gives only $4 \sim 5 \%$ contribution, because the process occurs 
at low momentum transfer, $q^2 = - 0.88 m_{\mu}^2$. 

Recently, Kirchbach and Riska \cite{Ki94} proposed a pseudovector 
form for the pion-induced component of the axial current, i.e. the second
term is replaced by ${  {g_P ( q^2)  } \over {  m_{\mu}}}
q^{\mu} ( \gamma \cdot q ) {\gamma}_5$ instead of including the last 
term in eq.(22). This procedure turns out to give the same result as 
that presented here for the spectrum of ${\mu}^{-} p \rightarrow
 n {\nu}_{\mu} \gamma$. On the other hand, Jenkins and Manohar \cite{Je91}
discussed on an effective Lagrangian which
contains higher derivative operators and the chiral symmetry breaking
quark mass matrix. However, in the present theory, this quark mass may be
absorbed into the pion effective mass.

As another attempt, the ChPT calculations of RMC have also been
carried out \cite{Me97,An97}, but the ${\hat g}_P$ = 6.77 value could 
not be extracted. Their results are, more or less, 
the same as those of Fearing's calculation \cite{Fe80,Be87}, 
in which the important ``Seagull'' term was missing. Thereby, these
calculations may have to be reexamined, taking higher order terms 
into account. It
should be noted that the ChPT can satisfy the gauge invariance
but it becomes obscure if the $\Delta$ degree of freedom
is taken into account.

In the present framework,
our calculation shows that ${\hat g}_P$ = 6.77 is
reasonable for both OMC and RMC on a proton.

{\bf Acknowledgment}

This work was supported by the KOSEF (961-0204-018-2) and the Korean Ministry
of Education (BSRI-97-2425). We thank F.Khanna for having called 
our attention to this problem.

\newpage

\centerline{\bf{FIGURE CAPTIONS}}

\vskip2cm

Fig. 1.$~$ Standard diagrams describing radiative muon capture on a proton.

Fig. 2.$~$ Photon energy spectrum for triplet states. 
The solid curve, which is to reproduce the experimental data reasonably, 
were taken from ref. \cite{Jo96} , i.e. the result without $\Delta M_e$
term for ${\hat g}_P$ = 9.8. The dotted curve is 
obtained without $\Delta M_e$ term for ${\hat g}_P = 6.77$. The dashed 
curve is with $\Delta M_e$ for ${\hat g}_P$ = 6.77. The dot-dashed curve
is calculated with $\Delta M_e$ term alone for ${\hat g}_P$ = 6.77.
(Figure of direct comparison with the experimental data is not
presented here because of some problems in PS file transform. Please
contact to the authors for more informations on our results)


\begin{references}
\bibitem{Ch97} It was firstly pointed out by the present author.
\end{references}
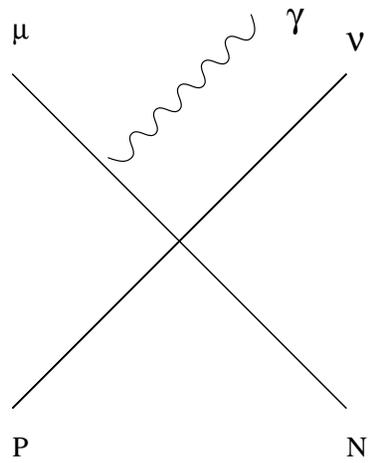
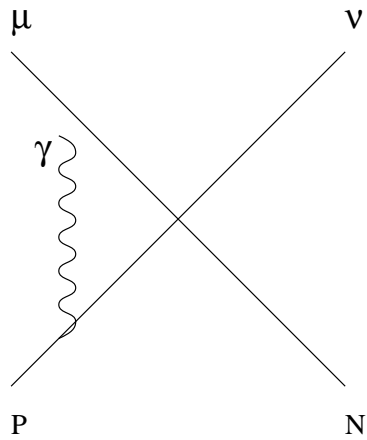
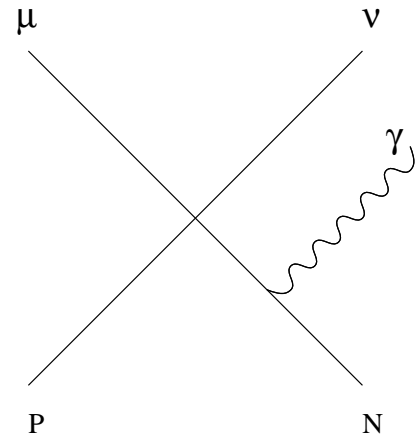
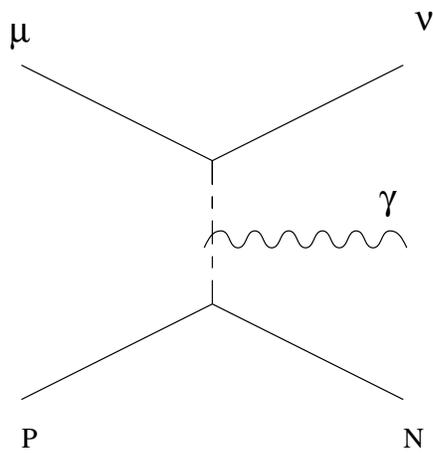
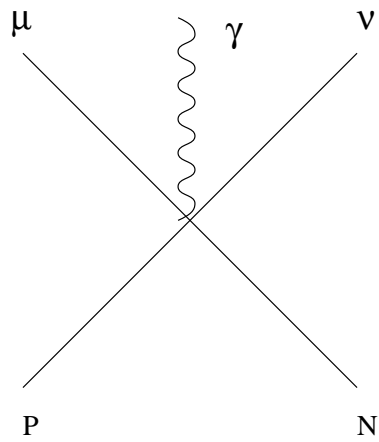
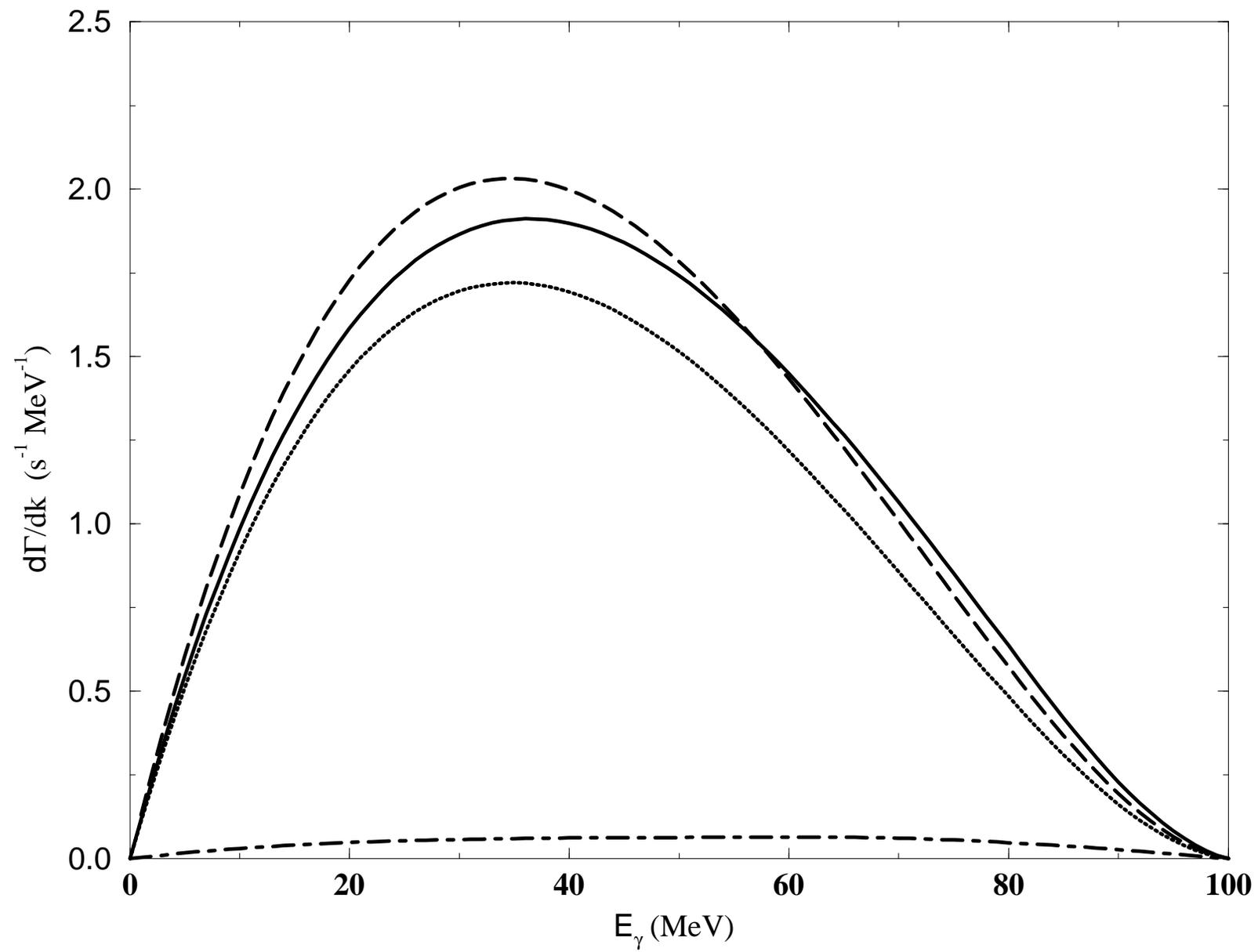
\end{document}